\documentclass[journal]{IEEEtran}
\usepackage{amsfonts}
\usepackage{amssymb}
\usepackage[cmex10]{amsmath}
\usepackage{graphicx}
\usepackage{subfigure}
\usepackage{cite}
\usepackage{tabularx,booktabs}
\usepackage{amsthm}
\usepackage{booktabs}
\usepackage[ruled,lined]{algorithm2e}
\usepackage{multirow}
\usepackage{setspace}
\usepackage{fancyhdr}
\usepackage{algorithmic}
\usepackage{enumerate}
\usepackage{color}
\usepackage{epstopdf}

\begin{document}
	\newcommand{\be}{\begin{equation}}
	\newcommand{\ee}{\end{equation}}
	\newcommand{\br}{{\mbox{\boldmath{$r$}}}}
	\newcommand{\bp}{{\mbox{\boldmath{$p$}}}}
	\newcommand{\bpi}{\mbox{\boldmath{ $\pi $}}}
	\newcommand{\bn}{{\mbox{\boldmath{$n$}}}}
	\newcommand{\balfa}{{\mbox{\boldmath{$\alpha$}}}}
	\newcommand{\ba}{\mbox{\boldmath{$a $}}}
	\newcommand{\bta}{\mbox{\boldmath{$\beta $}}}
	\newcommand{\bg}{\mbox{\boldmath{$g $}}}
	\newcommand{\bPsi}{\mbox{\boldmath{$\Psi $}}}
	\newcommand{\bsigma}{\mbox{\boldmath{ $\Sigma $}}}
	\newcommand{\bGamma}{{\bf \Gamma }}
	\newcommand{\bA}{{\bf A }}
	\newcommand{\bP}{{\bf P }}
	\newcommand{\bX}{{\bf X }}
	\newcommand{\bI}{{\bf I }}
	\newcommand{\bR}{{\bf R }}
	\newcommand{\bZ}{{\bf Z }}
	\newcommand{\bz}{{\bf z }}
	\newcommand{\bx}{{\mathbf{x}}}
	\newcommand{\bM}{{\bf M}}
	\newcommand{\bU}{{\bf U}}
	\newcommand{\bD}{{\bf D}}
	\newcommand{\bJ}{{\bf J}}
	\newcommand{\bH}{{\bf H}}
	\newcommand{\bK}{{\bf K}}
	\newcommand{\bm}{{\bf m}}
	\newcommand{\bN}{{\bf N}}
	\newcommand{\bC}{{\bf C}}
	\newcommand{\bL}{{\bf L}}
	\newcommand{\bF}{{\bf F}}
	\newcommand{\bv}{{\bf v}}
	\newcommand{\bSigma}{{\bf \Sigma}}
	\newcommand{\bS}{{\bf S}}
	\newcommand{\bs}{{\bf s}}
	\newcommand{\bO}{{\bf O}}
	\newcommand{\bQ}{{\bf Q}}
	\newcommand{\btr}{{\mbox{\boldmath{$tr$}}}}
	\newcommand{\bNSCM}{{\bf NSCM}}
	\newcommand{\barg}{{\bf arg}}
	\newcommand{\bmax}{{\bf max}}
	\newcommand{\test}{\mbox{$
			\begin{array}{c}
			\stackrel{ \stackrel{\textstyle H_1}{\textstyle >} } { \stackrel{\textstyle <}{\textstyle H_0} }
			\end{array}
			$}}
	\newcommand{\tabincell}[2]{\begin{tabular}{@{}#1@{}}#2\end{tabular}}
	\newtheorem{Def}{Definition}
	\newtheorem{Pro}{Proposition}
	\newtheorem{Lem}{Lemma}
	\newtheorem{Exa}{Example}
	\newtheorem{Rem}{Remark}
	\newtheorem{Cor}{Corollary}
	\newtheorem{The}{Theorem}
	\renewcommand{\labelitemi}{$\bullet$}

\title{Distributed multi-view multi-target tracking \\ based on CPHD filtering}

%

\author{{Guchong Li, Giorgio Battistelli, Luigi Chisci, Wei Yi, and Lingjiang Kong
	\thanks{This work was supported in part by the National Natural Science Foundation of China under Grant 61771110, in part by the Chang Jiang Scholars Program, in part by the 111 Project No. B17008.}		
	\thanks{G. Li, W. Yi, and L. Kong are with the School of Information and Communication Engineering, University of Electronic Science and Technology of China, Chengdu 611731, China, and G. Li is also with the Dipartimento di Ingegneria dell'Informazione, Universit\`{a} di Firenze, Florence 50139, Italy (e-mail: guchong.li@hotmail.com, kussoyi@gmail.com, lingjiang.kong@gmail.com).
		
	G. Battistelli and L. Chisci are with the Dipartimento di Ingegneria dell'Informazione, Universit\`{a} di Firenze, Florence 50139, Italy (e-mail: giorgio.battistelli@unifi.it, luigi.chisci@unifi.it).
}
}
}

\markboth{Journal of \LaTeX\ Class Files,~Vol.~X, No.~X, February~2020}%
{Shell \MakeLowercase{\textit{et al.}}: Bare Demo of IEEEtran.cls for IEEE Journals}

\maketitle

\begin{abstract}
	This paper addresses distributed multi-target tracking (DMTT) over a network of sensors having
different \emph{fields-of-view} (FoVs). 
Specifically, a \emph{cardinality probability hypothesis density} (CPHD) filter is run at each sensor node. 
Due to the fact that each sensor node has a limited FoV, the commonly adopted fusion methods become unreliable. 
In fact,  the monitored area of multiple sensor nodes consists of several parts that are either exclusive of a single node, i.e. \emph{exclusive FoVs} (eFoVs) or common
to multiple (at least two) nodes, i.e. \textit{common FoVs} (cFoVs).
In this setting, the crucial issue is how to account for this different information sets in the fusion rule.
The problem is particularly challenging when the knowledge of the FoVs is unreliable, for example
because of the presence of obstacles and target misdetection, or when the FoVs are time-varying.
Considering these issues, we propose an effective fusion algorithm for the case of unknown FoVs, where:
i) the intensity function is decomposed into multiple sub-intensities/groups by means of a clustering algorithm;
ii) the corresponding cardinality distribution is reconstructed by approximating the target \textit{random finite set} (RFS) as multi-Bernoulli; and iii) fusion is performed in parallel  according to either \emph{generalized covariance intersection} (GCI) or \emph{arithmetic average} (AA) rule. 
Simulation experiments are provided to demonstrate the effectiveness of the proposed approach. 
\end{abstract}


\section{Introduction}
\IEEEPARstart{S}{ensor} networks have undergone a rapid growth thanks to the development of  low-cost and low energy-consumption devices. One of the major tasks for such networks is 
\textit{distributed multi-target tracking} (DMTT), wherein sensor nodes perform local MTT and fusion by exchanging multi-target information with the other nodes. So far, DMTT has been successfully adopted in many fields such as environmental monitoring, air/ground/maritime surveillance, and sensor management \cite{Farina,BarShalom1988,Blackman1999,BarShalom2001,Mahler2014}.

Broadly speaking, the aim of MTT is the estimation of target number and states. 
A major problem challenge in achieving this objective is the unknown association of measurements to targets. 
In this respect, the recently developed \textit{random finite set} (RFS) approach \cite{Mahler2007,Mahler2014} has provided some effective solutions, 
such as the \textit{probability hypothesis density} (PHD) \cite{MahlerPHD} and \textit{cardinalized PHD} (CPHD) \cite{Mahler2007} filters, in which the target RFS is modeled as Poisson or, respectively, independent identically distributed (IID) cluster process. Instead of propagating the first-order moment and cardinality distribution like PHD and CPHD filters, another filtering approach, named \textit{multi-Bernoulli} (MB) filter \cite{Vo2009,Vo2010}, propagates the multi-target posterior density. 
Some effective implementations for these filters can be found by resorting to either the Gaussian Mixture (GM) \cite{Vo2006,Vo2007,Vo2009} or \textit{sequential Monte Carlo} (SMC) \cite{Sidenbladh2003,Vo2005,Vo2009} approaches.

Another key issue for DMTT is the selection of the fusion rule. 
Because of the unknown common information of sensor nodes, optimal fusion \cite{MahlerGCI2000} is ruled out. At the moment, there are two commonly adopted approaches for fusing multiple RFS densities: \textit{generalized covariance intersection} (GCI) \cite{Uney2013PHD,Giorgio2013CPHD,Battistelli2015,FVVBCrobust2018} and \textit{arithmetic average} (AA) \cite{GHBcauchy2017,GBCmultiobject2019,LCSpartial2018,LFGCsecond2019}.
\begin{enumerate}[$\bullet$]
	\item  \emph{GCI rule:} The GCI rule, which provides the weighted \textit{geometrical average} (GA) of the local RFS densities, is also named \textit{exponential mixture density} (EMD) in some papers \cite{Uney2013PHD}. From an information-theoretic point of view, it can be derived according to the \textit{minimum information gain} (MIG) criterion wherein the \textit{Kullback-Leibler divergence} (KLD) \cite{Kullback1951} is regarded as distance between densities \cite{Giorgio2013CPHD,GLkullback2014,Battistelli2015}. 
	It has been mathematically proved that GCI fusion can effectively avoid double-counting of common information.
Effective DMTT algorithms have been developed that combine 
GCI with several RFS-based (e.g., PHD, CPHD, multi-Bernoulli) filters,  \cite{Uney2013PHD,Giorgio2013CPHD,WYHLKYdistributed2017,LYLWKasynchronous2019,YLWHKcomputationally2020}. 
	It should be noted, however, that the GCI rule suffers from lack of robustness against high
 misdetection rates \cite{GBCmultiobject2019}.
	\item  \emph{AA rule:} The AA rule provides a weighted \textit{arithmetic average} (AA) of the local RFS densities, which also avoids the double-counting problem. 
From an information-theoretic viewpoint, AA  fusion can be derived according to the \textit{minimum information loss} (MIL) criterion based on KLD \cite{GBCmultiobject2019,DLZFFkullback2019}.
In fact, while AA fusion of Poisson processes  is no longer a Poisson process, the best fused Poisson process according to the MIL criterion can be obtained by arithmetically averaging the PHDs of the densities to be fused.
Similarly, for IID cluster processes, the best fused IID cluster process according to the MIL criterion can be
obtained by arithmetically averaging both cardinality distributions and PHDs.
	\cite{GBCmultiobject2019}. 
	Compared to the GCI rule, the AA rule shows its benefits or potential to alleviate the misdetection \cite{GBCmultiobject2019,GBCfusion2019} and cardinality inconsistency \cite{UHDJCfusion2018} problems. 
	DMTT algorithms combining RFS-based filters and AA fusion have been proposed in \cite{LCSpartial2018,GBCmultiobject2019,GBCfusion2019,LLPdistributed2020}. 	
\end{enumerate}

Although much progress has been made in DMTT, a relevant, not yet completely addressed, issue is that sensor nodes have limited \textit{fields-of-view} (FoVs), i.e. limited sensing range and/or angle. 
In such cases, how to extend the local FoVs to the overall FoV (union of all local FoVs) and achieve global tracking still remains a challenge. 
Unfortunately, the standard fusion rules (both GCI and AA) cannot satisfactorily deal with this objective. 
More specifically, GCI fusion tends to make targets outside the common FoV disappear because of its multiplicative (GA) nature, while
AA fusion preserves such targets at the expense of decreasing their weights. 
Motivated by this inadequacy of the existing fusion rules, some work, briefly summarized hereafter, has been developed to handle  multi-sensor fusion with multiple FoVs. 
\begin{enumerate}[$\bullet$]
	\item \emph{Within the GCI rule framework,} some simple but effective strategies, that consider the distance between \textit{Gaussian components} (GCs) or particles to fuse nearby targets while
	 keeping the targets that are in the exclusive FoV of only one node, are presented in \cite{GVMcooperative2016}. 
	 Moreover, a solution to handle multiple FoVs in the context of the \textit{simultaneous localization and mapping} (SLAM) is developed in \cite{BCLrandom2017}. 
	 Recently, a more principled method, where the  overall state space is decomposed into sub-spaces, is proposed in  \cite{GGWLdistributed2019}. 
	 Specifically, the method in \cite{GGWLdistributed2019} exploits  parallelized GCI fusion in the \emph{common FoVs} (cFoVs) and a compensation strategy in the \emph{exclusive FoVs} (eFoVs). 
	 Moreover, the issue of multiple FoVs  with GCI fusion has also been preliminarily investigated in the context of  labeled RFS-based filters \cite{LCCmultisensor2017,WGRXBHcentralized2018,LBCYWKmultisensor2018}.
	\item \emph{Within the AA rule framework,} a diffusion-based distributed SMC-PHD filter using sensors with limited sensing range has been presented in \cite{LEFClocal2019}. 
	Recently, AA fusion for labeled-RFS filters under multiple FoVs has been investigated in \cite{GBCfusion2019}.
\end{enumerate}

In general, for unlabeled RFS-based filters, the above mentioned approaches focus on the PHD filter. 
To the best of our knowledge, fusion of CPHDs under multiple different FoVs has not so far been addressed. 
In this paper, we tackle distributed CPHD-based filtering over a sensor network wherein nodes have different FoVs,
and either  GCI or AA fusion are employed in order to combine information among local CPHD filters. 
Summing up, the contributions of this work are the following.

\begin{enumerate}
	\item \emph{We demonstrate the impracticality of fusion under known FoVs.} 
	Although exact knowledge of the FoVs could easily accomplish integration of the local FoVs into the overall FoV, it is shown that the unavoidable uncertainties due to, e.g.,
	targets near the boundary of the FoV and misdetection of targets within the FoV caused by obstacles, can easily lead to overestimation or underestimation of the target number.  
	This analysis motivates investigation of the more realistic case of
	fusion under unknown FoVs.  
	\item \emph{We propose a robust distributed CPHD-based filter with unknown FoVs.} 
	In this context, adopting a GM implementation, the intensity functions to be fused are decomposed into sub-intensity functions
	by clustering the GCs.
	Then, to re-construct a CPHD form for each sub-intensity, the cardinality distribution is constructed by approximating the target RFS as multi-Bernoulli. 
	Next, the CPHD-based fusion between sub-intensities is carried out. After fusion, a new IID cluster process is constructed via convolution. The proposed solution is implemented using a GM approach.
	\item We show that the proposed clustering-based fusion strategy can be interpreted in terms of decomposition of the
	state space, and provide an analysis of the related approximation error. Simulation experiments are also given to demonstrate 	 	the effectiveness of the proposed approach.
\end{enumerate}

The outline of the rest of the paper is as follows. Section II introduces the background, and the problem statement is provided in Section III. The proposed fusion algorithm is described in Section IV. Simulation experiments are provided in Section V, and conclusions are drawn in Section VI.

\section{Background}
\label{sec:intro}
\subsection{Sensor network}
Suppose that $N$ sensor nodes make up a sensor network wherein each node can communicate with its neighbors, sense the environment and locally process the available data. 
Mathematically speaking, the network is represented by a graph $({\cal N}, \cal A)$, where ${\cal N}=\{1,2,\cdots,N\}$ denotes the set of nodes and ${\cal A} \subseteq {\cal N}\times {\cal N}$ the set of arcs (links). 
For any two nodes $i$ and $j$, there exists arc $(i,j)\in {\cal A}$ if node $j$ can receive data from node $i$.
Moreover, the in-neighborood of node $i$ is denoted as ${\cal N}^i=\{j\in{\cal N}:(j,i)\in {\cal A}\}$. 
The local FoV of node $l \in {\cal N}$, which represents the region wherein it can detect targets, is denoted as  ${\cal S}_l$.
Accordingly, ${\cal S} = \cup _{l = 1}^N{{\cal S}^l}$ is the overall FoV defined as the union of all local FoVs. 
Further, the cFoV between nodes $i$ and $j$ is defined as ${\digamma}_c^{i,j} = {{\cal S}^i} \cap {{\cal S}^j}$ and the eFoVs for nodes $i$ and $j$ are, respectively, denoted as $\digamma_{e}^{i,j} = {{\cal S}^i}\backslash \digamma_c^{i,j}$ and $\digamma_{e}^{j,i} = {{\cal S}^j}\backslash \digamma_c^{i,j}$.

\subsection{Multi-target Bayesian filter}

In random set tracking, targets and measurements, at time $k$, are regarded as RFSs
\begin{align}
\nonumber
X_k&=\{x_{k}^1,\cdots,x_{k}^{n_k}\} \in {\cal F}(\mathbb{X}), \\
\nonumber
Z_k&=\{z_{k}^{1},\cdots,z_{k}^{m_k}\}\in {\cal F}(\mathbb{Z}), 
\end{align}
where: 
$n_k$ and $m_k$ denote the unknown number of targets and, respectively, known number of measurements at time $k$; 
$\mathbb{X}$ and $\mathbb{Z}$ denote the state and, respectively, measurement space; 
$x_k^i \in \mathbb{X}$ and $z_k^j \in \mathbb{Z}$ are the $i$-th target state and, respectively, $j$-th measurement at time $k$; 
${\cal F}(\mathbb{X})$ denotes the set of finite subsets of $\mathbb{X}$.

Given the multi-target posterior density $f_{k - 1|k-1}(\cdot)$ at time $k-1$ and measurement set $Z_k$ at time $k$, the multi-target Bayes recursion can be expressed as follows
\begin{align}
\label{pre}
{f _{k|k - 1}}({X_k}) &= \int {{\varphi_{k|k - 1}}({X_k}|X){f _{k - 1|k-1}}(X) \,\delta X},\\
\label{upd}
{f _{k|k}}({X_k}) &= \frac{{{\ell_k}({Z_k}|{X_k}){f _{k|k - 1}}({X_k})}}{{\int {{\ell_k}({Z_k}|X){f _{k|k - 1}}(X) \,\delta X} }},
\end{align}
where $\varphi_{k|k-1}(\cdot|\cdot)$ and $\ell_k(\cdot|\cdot)$ are the \emph{multi-target transition density} describing the time evolution of the multi-target state and, respectively, the \emph{multi-object likelihood} describing the multi-target measurement model, respectively. The integrals in (\ref{pre}) and (\ref{upd}) are intended in the set integral sense \cite{Mahler2007}.

In general, the multi-target Bayesian filter can be hardly implemented unless the multi-target set $X_k$ is approximated, at each time $k$, by some
special RFS.
Hereafter, some typically used RFS approximations 
are summarized. 
%
\begin{enumerate}[$\bullet$]
	\item \emph{Multi-target Poisson RFS}: the multi-target posterior takes the form
	\begin{align}
{f_{k|k}}({X_k}) &= {e^{ - {\mu _{k|k}}}}\prod\limits_{i = 1}^n {{\mu _{k|k}} \, {s_{k|k}}({x_k^i})},
	\end{align}	
	where $\mu_{k|k}$ denotes the expected number of targets and $s_{k|k}$ is the target location density.
	Notice that all the information contained in a Poisson RFS can be summarized by its
	PHD (also called intensity function)
	\begin{equation}\label{eq:PHD} 
	v_{k|k} (x) = \mu_{k|k} \, s_{k|k} (x),
	\end{equation}
	which represents the first moment of the multi-target density. The integral of the PHD $v_{k|k}$ in any region of the state space corresponds
	to the expected number of targets contained in that region.

	\item \emph{IID cluster process}: the multi-target posterior takes the form	
\begin{align}
	{f_{k|k}}({X_k}) &= n!{p_{k|k}}(n)\prod\limits_{i = 1}^n {{s_{k|k}}({x_k^i})},
	\end{align}
	where $p_{k|k}$ is the cardinality distribution and  $s_{k|k}$ is again the target location density.
	The information contained in an IID cluster process can be summarized by the pair
	$\left (p_{k|k}, s_{k|k} \right)$ or, equivalently, by the pair  $\left (p_{k|k}, v_{k|k} \right)$
	where the PHD is as in (\ref{eq:PHD}) with
	\begin{equation}
	\mu_{k|k} = \sum_{n=0}^{\infty} n \, p_{k|k}(n).
	\end{equation}

	\item \emph{Multi-Bernoulli RFS}: the multi-target posterior takes the form
	\begin{align}
	\nonumber
	{f_{k|k}}({X_k}) &= n!\prod\limits_{i = 1}^n (1 - {r_i})\\
	&\times \sum\limits_{1 \le {i_1} \ne  \cdots  \ne {i_n} \le M} {\prod\limits_{j = 1}^n {\frac{{{r_{{i_j}}}{s_{{i_j}}}({x_j})}}{{1 - {r_{{i_j}}}}}} },
	\end{align}
	where $M$ is the number of Bernoulli components, $r_i$ and $s_i$ are the probability of existence
	and, respectively, location density of the $i$-th Bernoulli component.
\end{enumerate}

\subsection{Fusion rule}
Given multiple multi-target posterior densities, $\{\omega^i,f^i(X)\}$ with $\sum\nolimits_{i}\omega^i=1$, the
GCI and AA fusion rules will be considered.
In particular, assuming that the densities to be fused take the form of IID cluster processes $f^i(X) = n!p^i (n)\prod\limits_{x \in X} {s^i(x)}$, the GCI fusion provides  
\begin{align}
\label{gci_s}
\bar s_\text{GCI}(x) &= \frac{{\prod\limits_i {{{[{s^i}(x)]}^{{\omega^i}}}} }}{{\int {\prod\limits_i {{{[{s^i}(x)]}^{{\omega^i}}}} dx} }},\\
\label{gci_p}
\bar p_\text{GCI}(n) &= \frac{{\prod\limits_i {{{[{p^i}(n)]}^{{\omega^i}}}{{\left\{ {\int {\prod\limits_i {{{[{s^i}(x)]}^{{\omega^i}}}} dx} } \right\}}^n}} }}{{\sum\limits_{m = 0}^\infty  {\prod\limits_i {{{[{p ^i}(m)]}^{{\omega^i}}}{{\left\{ {\int {\prod\limits_i {{{[{s^i}(x)]}^{{\omega^i}}}} dx} } \right\}}^m}} } }}.
\end{align}
The AA fusion instead yields
\begin{align}
\label{aa_v}
\bar v_\text{AA}(x) &= \sum\limits_i {{\omega ^i}{v^i}(x)},\\
\label{aa_p}
\bar p_\text{AA}(n) &= \sum\limits_i {{\omega ^i}{p^i}(n)},
\end{align}
or, equivalently, in terms of target location density
\begin{equation}
\label{aa_s}
\bar s_\text{AA}(x) = \frac{1}{{\sum\limits_{n = 0} {n\bar p_\text{AA}(n)} }}\sum\limits_i {\left( {\sum\limits_{n = 0} {n{p^i}(n)} } \right){\omega^i}} {s^i}(x).
\end{equation}

There are essentially two ways to implement CPHD-based fusion, namely the GM approach \cite{Giorgio2013CPHD,GBCmultiobject2019} and, respectively, SMC (particle) approach \cite{Uney2013PHD}.

\section{Problem statement}
Although the GCI rule has been widely applied in DMTT scenarios, most of the papers assume that all nodes share the same FoV \cite{Giorgio2013CPHD,LGLP2019,WGRXHH2018}. 
In practice, however, different nodes turn out to have different FoVs. 
As a matter of fact, each sensor node can only detect targets over a portion of the area of interest and then exchanges information with the linked nodes (neighbors) aiming to extend 
the monitored area as much as possible, hopefully to the union of the FoVs of all sensor nodes. 
In this respect, fusion has to be handled with care, otherwise it could be counterproductive.
For instance,  because of its multiplicative nature [see (\ref{gci_s})], a direct application of the GCI fusion rule would tend to make targets outside the common FoV disappear.
Further, using AA fusion instead of GCI would only partially solve the problem since 
AA fusion tends to preserve all targets but with a reduced weight. In fact, the estimated number of targets after fusion corresponds to the AA of the estimated number of targets
of the local node, thus resulting in a severe underestimation when the FoVs overlap only partially.

To sum up, the fusion with multiple FoVs is still a thorny issue. 
Although PHD-based fusion with different FoVs has already been investigated \cite{LEFClocal2019,GGWLdistributed2019}
it is worth to point out that the PHD approach to DMTT, only keeping first-order moment information, may not be able to provide the desired tracking performance. 
In contrast, the CPHD approach might yield considerable performance improvements thus motivating our interest in this paper for the unaddressed topic of
CPHD-based fusion with different FoVs.

To make the fusion robust to multiple FoVs, based on our previous work in \cite{GGWLdistributed2019}, 
we resort to the idea of partitioning the 
global state space $\mathbb{X}$ into $G$ disjoint subspaces, i.e.,
\begin{align}
\mathbb{X} &= \mathbb{X}_1 \cup \mathbb{X}_2 \cup \cdots \cup \mathbb{X}_G,
\end{align}
\begin{align}
\mathbb{X}_g \cap \mathbb{X}_{g'} = \emptyset, ~~~~\forall g \neq {g'}.
\end{align}
Accordingly, the target set $X_k$ is partitioned as the union of $G$ disjoint RFSs
\begin{equation}
X_{k} =  {X}_{k,1} \cup {X}_{k,2} \cup \cdots \cup {X}_{k,G},
\end{equation}
with $X_{k,g} = X_k \cap \mathbb{X}_g$ representing the targets inside the subspace $\mathbb{X}_g$
at time $k$. For each $g$, we determine the set $\mathcal N_{k,g} \subseteq \mathcal N$ of the nodes of the
sensor network having information on the target subset $X_{k,g}$ at time $k$.
Then the idea is that for, each $g$, only the nodes belonging to $\mathcal N_{k,g}$ should be
involved in the fusion because they are the only ones providing information.

\subsection{Distributed CPHD filtering with different FoVs}

Suppose now that each sensor node is running locally a CPHD filter so that, after the local correction step,
the posterior density of each node $i$ takes the form of an IID cluster process
with cardinality distribution $p_{k|k}^i$ and target location density $s_{k|k}^i$. Notice that,
in order to simplify the notations, hereafter we will drop the time dependence and simply write $p^i$ and $s^i$
where it is understood that all quantities refer to the current time instant $k$.
The key steps in applying the  above-described idea to the fusion of IID cluster processes are:
\begin{enumerate}[a)]
\item for each node $i$, split the original IID cluster density $(p^i, s^i)$ into suitable sub-IID cluster densities
$(p^i_g, s^i_g)$ each one corresponding to one of the subspaces $\mathbb X_g$; 
\item
 for any $g$, carry out fusion between the sub-IID cluster densities $(p^i_g, s^i_g)$ 
with $i \in \mathcal N_{k,g}$ to get the fused sub-IID cluster density $(\overline p_g, \overline s_g)$;
\item after fusion, combine the  fused sub-IID cluster density $(\overline p_g, \overline s_g)$ with $g = 1, \ldots, G$
to get the total fused IID cluster density $(\overline p, \overline s)$.
\end{enumerate}

Notice that in the above procedure, each  target subset $X_{k,g}$ is modelled as an IID cluster process.
Accordingly, the total target set $X_k$ is modelled as a superposition (independent union) of IID cluster processes.
This means that, given the fused cardinality distributions $\overline p_g$ and PHD $\overline v_g$ pertaining to
each subset $ X_{k,g}$, the total fused cardinality distribution $\overline p$ and PHD $\overline v$ in step c) can
be readily obtained as
\begin{eqnarray*}
\overline p(n) &=& (\overline p_1 * \cdots * \overline p_G) (n), \\
\overline v(x) &=& \overline v_1 (x) + \ldots + \overline v_n (x),
\end{eqnarray*} 
where $*$ stands for discrete convolution. 

The converse operation, that is the splitting of the total  IID cluster density $(p^i, s^i)$ into the  sub-IID cluster densities $(p^i_g, s^i_g)$ which has to be performed in step
a), can be carried out by resorting to a 
multi-Bernoulli approximation as proposed in \cite[section 9.4.2]{Mahler2014}.

First of all, notice that the total intensity $v^i$ can be readily splitted as
\begin{equation}
v^i (x) = v^i_1 (x) + \ldots + v^i_G (x),
\end{equation}
where each sub-intensity is obtained as
\begin{equation}\label{eq:split}
 v^i_g (x) = 1_{\mathbb X_g} (x) \, v^i (x),
\end{equation}
$ 1_{\mathbb X_g} (x)$ denoting the indicator function of set $\mathbb X_g$.

Let us assume now, without loss of generality, that each sub-intensity can be expressed as
\begin{align}\label{decom_PHD}
v^i_g (x)= \sum\limits_{q=1}^{J_g^i}\alpha^i_{g,q} f^i_{g,q}(x),
\end{align}
where $\alpha^i_{g,q} \in (0,1)$ and $\int_{\mathbb X} f^i_{g,q}(x) d x=1$.
In this way, the existence probabilities $\alpha^i_{g,q}$ and spatial probability density functions (PDFs) 
$ f^i_{g,q}$ provide the best multi-Bernoulli approximation.
Then, exploiting  \cite[eqn. (9.29)]{Mahler2014},
the cardinality distribution of the $g$-th sub-IID cluster process can be computed as follows
\begin{equation}
{p_g^i}({n}) = \prod\limits_{j = 1}^{{J_g^i}} {(1 - {\alpha^i_{g,q}})} {\sigma _{{J_g^i},{n}}}\left( \frac{\alpha^i_{g,1}}{1 - \alpha^i_{g,1}}, \cdots ,\frac{{{\alpha^i_{g,{J^i_g}}}}}{{1 - {\alpha^i_{g,{J^i_g}}}}} \right),
\end{equation}
where
\begin{align}
{\sigma _{m,n}}({\beta_1}, \cdots ,{\beta_m}) = \left\{ {\begin{array}{*{20}{cl}}
	1 &  \mbox{if } n = 0\\
	\sum\limits_{1 \le {i_1} <  \cdots  < {i_n} \le m} {\prod\limits_{j = 1}^n {{\beta_{{i_j}}}} } & \mbox{if } 1 \le n \le m\\
	0 & \mbox{if } i > m
	\end{array}} \right.
\end{align}
is called \emph{elementary symmetric function} (ESF) of degree $n$ in $m$ variables.

\subsection{On the decomposition of the state space}

The fusion procedure described in the previous section presupposes that the state space is already
partitioned into disjoints sets $\mathbb X_1, \ldots, \mathbb X_G$ and that, for each $\mathbb X_g$,
we know which of the sensor nodes have information on the targets belonging to   $\mathbb X_g$
(i.e., for each $g$ the set $\mathcal N_{k,g}$ is supposed to be known).

\begin{figure}[tb]
	\centering
	\includegraphics[width=1.0\columnwidth,draft=false]{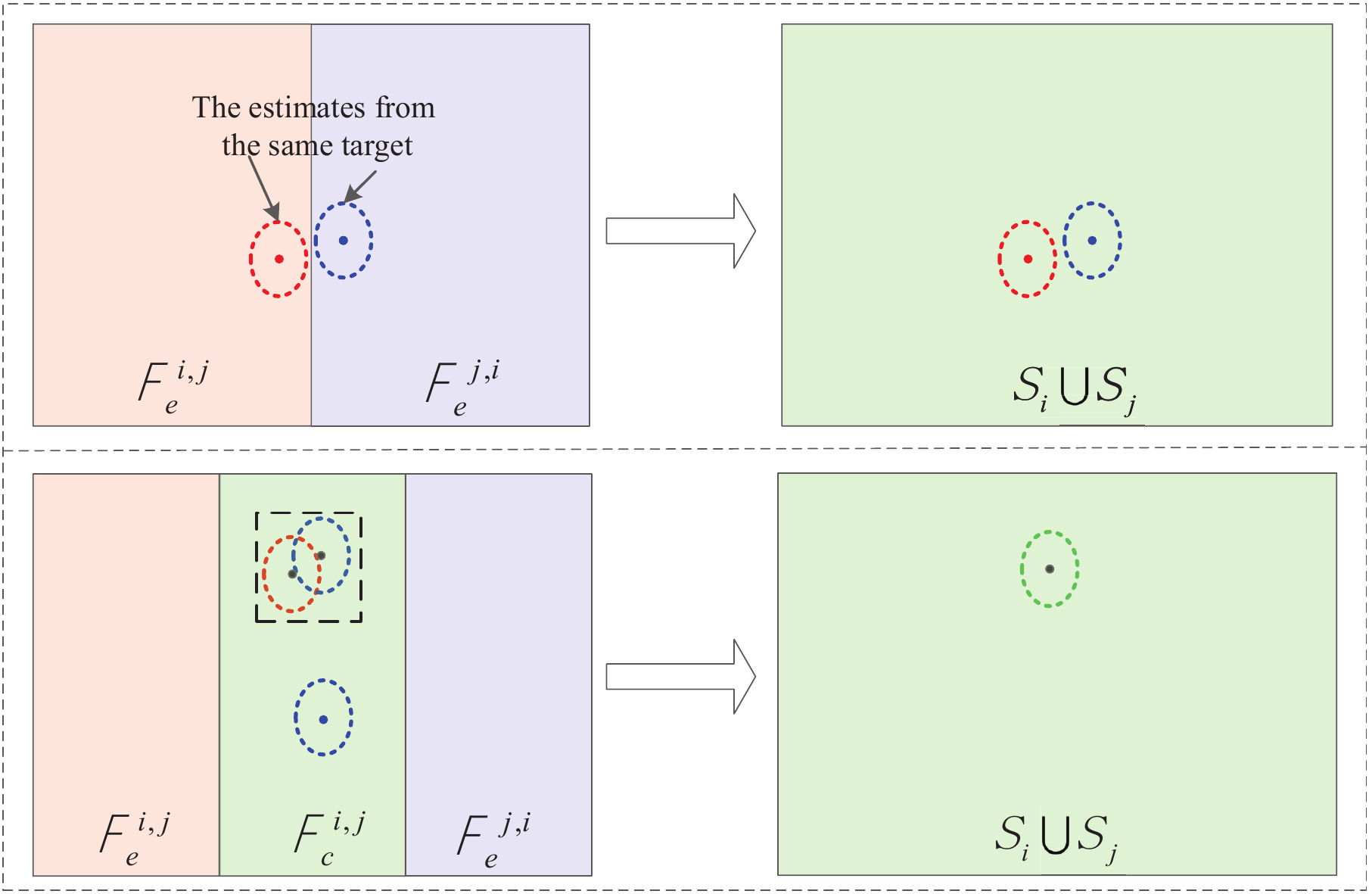}
	\caption{(a) Two GCs, actually representing the same target, are located in the eFoVs of the two fusing nodes $i$ and $j$. 
	Hence, they are both copied in the fused density thus implying overestimation of the target number. 
	(b) Three GCs are located in the cFoV of nodes $i$ and $j$; two of them, corresponding to the same target, are merged after fusion while the third one, corresponding to a target detected by node $j$ but node by node $i$, disappears after fusion thus implying underestimaton of the target number.
	} 
	\label{fig:problem}
\end{figure}

If the FoVs were precisely known,
a possible \textit{common-sense} strategy 
could be to define the state space partition on the basis of the FoVs so as to 
fuse information within the cFoV while copying information from the eFoVs.
Unfortunately, such common-sense fusion strategy lacks of robustness with respect to the unavoidable uncertainties on the FoVs
as it can be seen from the two examples illustrated in Fig. \ref{fig:problem}, concerning fusion between two nodes.

\begin{enumerate}[$\bullet$]
	\item \emph{Targets located in the eFoV.}\\ 
	In the example of Fig. \ref{fig:problem}(a), the local density $f^i(\cdot)$ of node $i$ has a 
Gaussian Component (GC) located in the eFoV $\digamma_{e}^{i,j}$ and similarly
	for node $j$ in $\digamma_{e}^{j,i}$.	
	Due to their closeness, the two GCs clearly correspond to the same target.
	Then, applying common-sense fusion, they are both copied into the fused density thus implying an overestimation of the target number.
	\item \emph{Targets located in the cFoV.}\\ 
	In the example of Fig. \ref{fig:problem}(b), the local density $f^i(\cdot)$ of node $i$ has a single GC located in the cFoV 
	${\digamma}_c^{i,j}$ while for node $j$ there are two GCs in ${\digamma}_c^{i,j}$.
	Notice that two GCs correspond to the same target and, after fusion, are merged into a single one while the other GC, which corresponds to a target detected by node $j$ but undetected by node $i$, tends to be removed after fusion. In this case, underestimation of the target number occurs.	
\end{enumerate}

Further, the common-sense startegy cannot be applied when the FoVs are not perfectly known
(for example due to occlusions) or when the sensors have information also outside their current FoVs
(for example because the FoVs are time-varying or because of the fusion feedback at previous time instants). 

For the above reasons, in this paper we give up on defining the state-space partition on the basis of the
sensor FoVs and, instead, propose a procedure that exploit clustering to directly 
 split the original IID cluster density into suitable sub-IID cluster densities and, then, carry out fusion
without knowledge of the FoVs.


\section{Robust distributed CPHD filtering with multiple FoVs}

To make the CPHD-based fusion immune to multiple FoVs, a robust fusion strategy based on clustering is now proposed. Section IV-A provides a detailed description of the fusion algorithm, while Section IV-B 
 shows how the proposed approach is related to the decomposition of the state space introduced in Section II.
Without loss of generality, pairwise fusion between two nodes, say $1$ and $2\in{\cal N}^1$, will be considered but the resulting fusion approach will admit a straightforward extension
to an arbitrary number of fusing nodes, i.e., it can be applied simultaneously to any subset of $\mathcal{N}$.

\subsection{Robust CPHD-based fusion with unknown FoVs} 

Following a GM implementation, it is assumed that the local intensities of the sensor nodes are
expressed as 
\begin{equation}
v^i (x) = \sum_{p=1}^{J^i} \alpha_p^i \, {\cal G}(x;m_p^i,P_p^i)
\end{equation}
for $i \in \{ 1, 2 \}$, where $\alpha_p^i \in (0, 1)$ and ${\cal G}(x;m,P)$ denotes a Gaussian PDF with mean $m$ and covariance $P$.   
The proposed fusion algorithm consists of four steps detailed hereafter.
\subsubsection{Clustering of GCs}
In order to determine whether two generic GCs need to be grouped into the same cluster or not, the idea is to compare
a suitable distance between the GCs to an appropriately chosen threshold.
Specifically, let ${\cal G}^i_p$ and ${\cal G}^j_{q}$
be two generic GCs\footnote{For convenience, the GC, ${\cal G}(x;m^i_p,P^i_p)$, is simply denoted as ${\cal G}^i_p$, and similarly for ${\cal G}^j_{q}$.} with $i, j \in \{1, 2\}$. 
Then, their distance is defined in terms of the \emph{corrected Mahalanobis distance},
\begin{equation}\label{eq:Mah}
{d({\cal G}^i_p,{\cal G}^j_{q})} = {({m_p^i} - {m_{q}^j})^{\top}}{({P_p^i} + {P_{q}^j})^{ - 1}}({m_p^i} - {m_{q}^j}),
\end{equation}
If ${d({\cal G}_p^i,{\cal G}_{q}^j)}$ is smaller than a preset gating threshold $\rho$, then ${\cal G}_p^i$ and ${\cal G}_{q}^j$ are assigned to the same cluster. 
It is clear that the farther apart the two GCs are, the less similar they are. 
As a matter of fact, a large distance makes the common information between GCs negligible. 
As a consequence, the intensity function can be approximately split into sub-intensity functions corresponding to the obtained clusters. 
The detailed clustering process is provided in \emph{Algorithm 1}.
\begin{algorithm}[htb]
	\DontPrintSemicolon
	\caption{{GC clustering}}
	\LinesNumbered
	\small\underline{\sc {INPUT}}: GC set $\{{\cal G}_p^i\}_{p=1}^{J^i}$ and index set
	${\cal I}^i = \{ (i,p) \}_{p=1}^{J^i}$ for $i = 1,2$;
	 gating threshold $\rho$.\;
	\small\underline{\sc 
		{{OUTPUT}}}: Clustering set ${\cal C}$.\;
	\textbf{Function} {Clustering$\_$Function} \\
	INITIALIZE the index set as ${\cal I}={\cal I}^1\cup{\cal I}^2$;\\
	INITIALIZE the number of clusters as $g=0$;\\
	\While{non-empty ($\cal I$)}{
		Select a GC index $(i,p) \in {\cal I}$; \;
		$g =: g+1$; \;
		INITIALIZE  the $g^{th}$ cluster as ${\cal C}_g  = \emptyset$; \;
		\For{ $(j,q) \in {\cal I}$}{
			Compute the distance $d({\cal G}_p^i,{\cal G}_q^j)$; \;
			\If{$d({\cal G}_p^i,{\cal G}_q^j)<\rho$}{
				${\cal C}_g ={\cal C}_g  \cup \{ (j,q)\}$; \;
			}  	
		}
		Delete the indices of used GCs, i.e. ${\cal I} = {\cal I}\backslash {\cal C}_g$. \;
	}
	$G=g$; \;
	\textbf{Return:} Clustering set ${\cal C}=\{{\cal C}_g\}_{g=1}^G$.
\end{algorithm}


Based on the clustering process, a feasible definition of the \emph{clustering set} is given as follows.
\begin{Def}
A set ${\cal C} = \{{\cal C}_1,\cdots,{\cal C}_G\}$ with $G$ clusters of GC indices is
 a \textit{clustering set} if the following two conditions hold:
 \begin{enumerate}[$\blacktriangleleft$]
		\item for any cluster ${\cal C}_g$ and for any $(i,p) \in {\cal C}_g$,
			there exists $(j,q) \in {\cal C}_g$ such that
		\begin{align}\label{def1}
		d({\cal{G}}_p^i,{\cal{G}}_q^j) \le \rho \, ;
		\end{align}
		\item for any $(i,p) \in {\cal C}_g $ and $(j,q) \in {\cal C}_{g'} $ with $g \ne g'$
			then 
		\begin{align}\label{def2}
		d({\cal{G}}_p^i,{\cal{G}}_q^j) > \rho
		\end{align}
	\end{enumerate}
where $\rho$ is the predefined gating threshold.
\end{Def}

Note that (\ref{def1}) is automatically satisfied according to the distance thresholding of Algorithm 1, while (\ref{def2}) may not be satisfied.
In order to ensure also (\ref{def2}), we resort to the so-called \textit{union-find set} (UFS) algorithm so that the distance between any two GCs from different clusters is larger than 
$\rho$, meaning that the clusters are  well-separated.

As a result of clustering, the intensities of nodes $1$ and $2$ can be decomposed as
\begin{equation}
v^i (x) = \sum_{g=1}^G \hat v_g^i (x)
\end{equation}
for $i \in \{ 1, 2 \}$ where each sub-intensity $\hat v_g^i (x)$ contains the GCs
whose index $(i,p)$ belongs to ${\cal C}_g$.
In particular, let $J_g^i$ denote the number of GCs of node $i$ belonging to cluster ${\cal C}_g$.
Then, when $J_g^i >0$ the corresponding sub-intensity can be represented as 
\begin{align}
{{\hat v}_g^i}(x) &= \sum\limits_{p = 1}^{{J_g^i}} {\alpha_{g,p}^i \, {\cal G}(x;m_{g,p}^i,P_{g,p}^i)}.
\end{align}
Conversely, when $J_g^i =0$ we simply have ${{\hat v}_g^i}(x) = 0$. This latter case
corresponds to the situation in which node $i$ does not have any GC in cluster ${\cal C}_g$.
In the proposed fusion strategy, the sub-intensities $\hat v_g^i$ obtained through clustering
are used in place of the sub-intensities $v_g^i$ obtained according to the partitioning of
the state space as in (\ref{eq:split}). Then, the set $\mathcal N_{k,g}$ is defined as
$\mathcal N_{k,g} = \{ i: J_{g}^i > 0\}$ so that, for each cluster ${\cal C}_g$, only the sensor nodes having GCs
belonging to such a cluster are involved in the fusion.

\begin{Rem}
	It is clear that clustering highly depends on the selection of the threshold $\rho$. 
	\begin{enumerate} [-]
		\item Large $\rho$ might lead to group GCs from different targets into the same cluster. Consider two GCs from node $1$, ${\cal G}_1^1 \in \digamma_c^{1,2}$ and ${\cal G}_2^1 \in \digamma_{e}^{1,2}$, and one GC from node $2$,  ${\cal G}_1^2 \in \digamma_c^{1,2}$, where ${\cal G}_1^1$ and ${\cal G}_1^2$ are from the same target. 
		If the three GCs are grouped into only one cluster, 
then the target represented by ${\cal G}_2^2$ will not be matched by any GC of node $1$ so that the number of targets will be underestimated after fusion.
		\item Small $\rho$ may enforce the GCs from the same target to be grouped into different clusters. 
		For instance, if $v^1(x)=0.9{\cal G}_1^1  \in \digamma_c^{1,2}$ and $v^2(x)= {\cal G}_1^2 \in \digamma_c^{1,2}$ with 
		$d \left( {\cal G}_1^1, {\cal G}_1^2 \right) > \rho$, it can happen that these GCs are grouped into two different clusters even if they correspond to the same target. In such a case, the number of targets will be overestimated. 
	\end{enumerate} 
Hence, the selection of $\rho$ is a trade-off between underestimation and overestimation. 
For scenarios with well-separated targets, $\rho$ can be easily chosen.
A more formal analysis of the role played by the gating threshold $\rho$ will be given in Section IV-B.

\end{Rem}

\subsubsection{Construction of the cardinality distribution of each cluster}
For the CPHD filter, one major advantage is the use of the cardinality distribution to estimate the number of targets according to the \textit{maximum a posteriori probability} (MAP) criterion
rather than the \textit{expected a posteriori} (EAP) criterion adopted by the PHD filter. 
Hence, the construction of the cardinality distribution is needed. 

For the obtained $g$-th cluster and the corresponding sub-intensities ${\hat v}_g^i(x)$, $i \in \{1, 2\}$, 
by using the Bernoulli approximation, the cardinality distributions can be obtained as follows:
\begin{align}
\nonumber
&\hat p_g^i(n) \\
&= \left( {\prod\limits_{p = 1}^{J_g^i} {(1 - {\alpha_{g,p}^i})} } \right){\sigma _{{J_g^i},n}}\left( {\frac{{{\alpha_{g,1}^i}}}{{1 - {\alpha_{g,1}^i}}}, \cdots ,\frac{{{\alpha_{g,{J_g^i}}^i}}}{{1 - {\alpha_{g,{J_g^i}}^i}}}} \right)
\end{align}
for $i \in \{ 1, 2  \}$.
\subsubsection{Fusion for each cluster}
For any cluster ${\cal C}_g$, $g\in \{1,\cdots,G\}$, there are two possible cases depending on the origin of the GCs within the cluster.
\begin{enumerate}[-]
	\item all GCs belong to a single node $i$ (i.e. $\mathcal N_{k,g} = \{ i \}$): 
	keep the sub-intensity and the cardinality distribution of such a node
	\begin{align}
	\overline v_g (x) &= \hat v_g^i (x), \\
	\overline p_g (x) &= \hat p_g^i (x).
	\end{align}
	\item GCs belong to both nodes $1$ and $2$  (i.e. $\mathcal N_{k,g} = \{ 1,2 \}$):  
	$({\hat{v}_g^1},\hat p_g^1)$ and $({\hat{v}_g^2},\hat p_g^2)$ need to be fused 
	by using either the GCI or the AA fusion rule. \\
	\underline{GCI fusion}
	The fused target location density for cluster ${\cal C}_g$ can be obtained by resorting to any
	of the available techniques for computing the GA of GMs. For example, when the approximation of 
	\cite{Giorgio2013CPHD} is adopted, the fused target location density takes the form
	\begin{align}
	\label{gci_gm}
	\overline s_{g}(x) = {C^{ - 1}} \, \sum\limits_{p = 1}^{{J_g^1}} {\sum\limits_{q = 1}^{{J_g^2}} {\alpha _{pq}^{1,2} \, {\cal G}(x;m_{pq}^{1,2},P_{pq}^{1,2})} }
	\end{align}
	where 
	\begin{align}
	\nonumber
	P_{pq}^{1,2} =& {[\omega {(P_{g,p}^1)^{ - 1}} + (1 - \omega ){(P_{g,q}^2)^{ - 1}}]^{ - 1}}, \\
	\nonumber
	m_{pq}^{1,2} =& P_{pq}^{1,2}[\omega {(P_{g,p}^1)^{ - 1}}m_{g,p}^1 + (1 - \omega ){(P_{g,q}^2)^{ - 1}}m_{g,q}^2], \\
	\nonumber
	\alpha _{pq}^{1,2} =& {(\alpha _{g,p}^1)^\omega }{(\alpha _{g,q}^2)^{1 - \omega }}\kappa (\omega ,P_{g,p}^1)\kappa (1 - \omega ,P_{g,q}^2)\\
	\nonumber
	&\cdot{\cal G}(m_{g,p}^1 - m_{g,q}^2;0,{\textstyle{{P_{g,p}^1} \over \omega }} + {\textstyle{{P_{g,q}^2} \over {1 - \omega }}})
	\end{align}
	and
	\begin{align}
	C &= \sum\limits_{p = 1}^{{J_g^1}} {\sum\limits_{q = 1}^{{J_g^2}} {\alpha _{pq}^{1,2}} },\\
	\kappa (\omega ,P) &= \frac{{{{[\det (2\pi P{\omega ^{ - 1}})]}^{{\textstyle{1 \over 2}}}}}}{{{{[\det (2\pi P)]}^{{\textstyle{\omega  \over 2}}}}}}.
	\end{align}
	Substituting (\ref{gci_gm}) into (\ref{gci_p}), the cardinality distribution $\overline{p}_{g}(n)$ can be obtained. Moreover, the fused sub-intensity function can be expressed as 
	follows:
	\begin{align}
	\overline v_{g}(x) = \overline s_{g}(x)\sum\limits_{{n} \ge 0} {{n} \, \overline p_{g}({n})}.
	\end{align}
	\underline{AA fusion}
	The fused sub-intensity function and cardinality distribution can be readily computed without any approximation as follows
	\begin{align}
	\nonumber
	{\overline v_{g}}(x) &=  \omega \sum\limits_{p = 1}^{{J_{g}^1}} {{\alpha _{g,p}^1} \, {\cal G}(x;m_{g,p}^1,P_{g,p}^1)} \\
	&{\kern 8pt}+ (1 - \omega )\sum\limits_{q = 0}^{{J_{g}^2}} {{\alpha _{g,q}^2} \, {\cal G}(x;m_{g,q}^2,P_{g,q}^2)}, \\
	{\overline p}_{g}(n) &= \omega \, \hat p_g^1(n) + (1-\omega) \, \hat p_g^2 (n).
	\end{align}
\end{enumerate}

\subsubsection{Merging sub-IID cluster processes}

Once the cluster-wise fusion has been carried out, all fused sub-IID cluster processes should be merged into a single IID cluster process.
The merging is obtained by summing all sub-intensity functions and convolving the corresponding cardinality distributions as follows:
\begin{align}
\overline v(x) =& \sum\limits_{g = 1}^G {{\overline v_{g}}(x)},\\
\nonumber
\overline p^i(n) =& \,( {\overline p_{1}}*{\overline p_{2}}*\cdots*{\overline p_{G}} )(n)  \vspace{1mm} \\
=& \sum\limits_{{n_1} + {n_2} +  \cdots  + {n_G}=n} {{\overline p_{1}}({n_1}) \cdot } {\overline p_{2}}({n_2}) \cdots {\overline p_{G}}({n_G}).
\end{align} 

\begin{Rem}
	It is worth to point out that clustering does not involve the weights of the GCs. That is to say, the focus is only on the spatial positions of targets but not on the existence probabilities. The following two cases need to be emphasized.
	\begin{enumerate}[-]
		\item If one target travels only in the FoV of one of the two fusing nodes, no fusion is performed and the GCs of the target will be preserved so that the target can be detected by both nodes. In such a case, both AA and GCI fusion will provide good cardinality estimation.
		\item Under low detection probability, it may happen that in the cFoV there are two GCs from different nodes in  the same cluster with comparatively different weights, i.e.
		one very small and the other greater than $0.5$.
		In such a case, the target estimation accuracy of the fused GC, no matter whether GCI or AA fusion is adopted, can be improved but the weight of the fused GC may decrease compared to the initial GC with larger weight. Hence, cardinality may be underestimated. It should be noticed that in a sensor network, it is difficult to guarantee that all nodes can simultaneously detect all the targets. 
		In this respect, GCI fusion, due to its multiplicative nature, is more sensitive to target misdetection than AA fusion which, on the other hand, has additive nature. 
	\end{enumerate}
\end{Rem}

\begin{Rem}
	Although the above algorithmic description concerns only two sensor nodes (pairwise fusion), it is easy to extend it to an arbitrary number of nodes, e.g., 
	by means of a sequence of pairwise fusions. 
\end{Rem}
\begin{Rem}
In alternative to the previously considered GM implementation, also a SMC (particle) approach, using delta-function mixtures instead of GMs, could be adopted.
In such a case,
the Euclidean distance can be exploited in order to measure the closeness between particles in the clustering process.
Further, the \textit{Kernel Density Estimation} (KDE) method \cite{BSdensity1986}  could be used 
for GCI fusion of delta-function mixtures.
\end{Rem}
\begin{Rem}
Notice that the proposed clustering-based fusion strategy is well-suited to being applied also in the context
of distributed PHD filtering with different FoVs. In this case, clearly, the operations related to splitting and fusion of the
cardinality distribution need not be performed.
\end{Rem}

\subsection{Approximation error analysis}

In this subsection, we provide an interpretation of the proposed clustering-based fusion strategy in terms of decomposition of the state
space, and analyze the error induced by approximating the sub-intensities $v_g^i (x)$ with the GMs $\hat v_g^i (x)$.

To this end, given a GC ${\cal G}^i_p$ and a scalar $\delta$, let us consider the confidence ellipsoid
\begin{equation}
{\cal E}^i_p (\delta) = \left \{ x \in \mathbb X:\; (x-m_p^i)^\top (P_p^i)^{-1} (x-m_p) \le \delta \right \}.
\end{equation}
Recall that, after step 1) of the proposed strategy, the index partition ${\cal C} = \left \{ {\cal C}_1, \ldots, {\cal C}_g \right \}$
defines a clustering set satisfying both properties in Definition 1. As a consequence the following result holds.

\begin{Lem} Let $\delta$ be such that $\delta \le \rho/4$, $\rho$ being the gating threshold. Then, for any
$(i,p) \in {\cal C}_g$ and $(j,q) \in {\cal C}_{g'}$ with $g \ne g'$, we have
\begin{equation}
{\cal E}^i_p (\delta) \cap {\cal E}^j_q (\delta) = \emptyset.
\end{equation}
\end{Lem}

\noindent
\textit{Proof:} see Appendix A. \qed

In turn, Lemma 1 implies that we can partition the state space $\mathbb X$ into $G$ disjoint sets $\mathbb X_1 , \ldots, \mathbb X_G$
having the properties that
\begin{align}
(i,p) \in {\cal C}_g & \implies {\cal E}^i_p (\delta)  \subseteq \mathbb X_g,  \label{eq:prop1} \\
(i,p) \notin {\cal C}_g & \implies {\cal E}^i_p (\delta)  \cap \mathbb X_g = \emptyset. \label{eq:prop2}
\end{align}
Hence the clustering procedure based on the corrected Mahalanobis distance (\ref{eq:Mah}) implicitly
defines a partitioning of the state space.

Consider now the sub-intensities $v_g^i (x)$ determined by splitting each $v^i (x)$ with respect to the above-defined
partition
\begin{equation}
v^i_g (x) = 1_{\mathbb X_g} (x) \, v^i (x) = \sum_{p = 1}^{J^i} \alpha_p^i \, 1_{\mathbb X_g} (x)  \, {\cal G}^i_p (x),
\end{equation}
and the corresponding sub-intensities $\hat v^i_g (x)$ obtained by means of the proposed clustering procedure
\begin{equation}
\hat v^i_g (x) = \sum_{p: \, (i,p) \in {\cal C}_g } \alpha_p^i \,  {\cal G}^i_p (x).
\end{equation}
Notice that the discrepancy between each pair of sub-intensities can be written as
\begin{align}
v^i_g (x) - \hat v^i_g (x) =&  \sum_{p: \, (i,p) \in {\cal C}_g } \alpha_p^i \, \left [ 1- 1_{\mathbb X_g} (x)\right ] \, {\cal G}^i_p (x)
\nonumber \\
& + \sum_{p: \, (i,p) \notin {\cal C}_g }   \alpha_p^i \, 1_{\mathbb X_g} (x)  \, {\cal G}^i_p (x). \label{eq:error}
\end{align}
Recall now that for a Gaussian PDF ${\cal G}^i_p$ we have
\begin{equation}
\int_{{\cal E}^i_p (\delta)} {\cal G}^i_p (x)\,  d x = F (\delta , \dim (x)),
\end{equation}
where $F(\delta, d)$ denotes the cumulative distribution function of the $\chi^2$-distribution with $d$ degrees of freedom.
Then, by exploiting properties (\ref{eq:prop1})-(\ref{eq:prop2}), a bound on the approximation error
(\ref{eq:error}) in $L_1$-norm can be obtained.

\begin{The}
 Let $\delta$ be such that $\delta \le \rho/4$ with $\rho$ the gating threshold, and let 
$\mathbb X_1 , \ldots, \mathbb X_G$ be any partition of the state space $\mathbb X$ satisfying properties
  (\ref{eq:prop1})-(\ref{eq:prop2}). Then, the approximation error between 
$v_g^i (x)$ and $\hat v_g^i (x)$ can be upper-bounded as follows
\begin{equation}
\left \| v_g^i (x) - \hat v_g^i (x) \right \|_1 \le \mu^i \, \left [ 1- F(\delta , \dim (x))\right ],
\end{equation}
where $\mu^i$ is the estimated number of targets
\begin{equation}
\mu^i = \sum_{p=1}^{J^i} \alpha_p^i.
\end{equation}
\end{The}
\noindent
\textit{Proof:} see Appendix A. \qed

Notice that the approximation error tends to vanish as $\delta$ (and hence $\rho$) increases, because
the portion of each Gaussian ${\cal G}^i_p$ with $(i,p) \in {\cal C}_g$ contained in the corresponding
subset $\mathbb X_g$ increases. 
In fact, a large $\rho$ implies that GCs belonging to different clusters are well separated.
In the limiting case, the approximation error can be neglected if all clusters are well-separated.
On the other hand, when $\rho$ is large  a sub-intensity function may not necessarily correspond to only one target, but might also represent multiple closely-spaced targets.

\section{Performance evaluation}
In this section, the proposed robust distributed fusion with unknown multiple FoVs is applied to both CPHD and PHD filters and the resulting DMTT performance is evaluated by means of simulation experiments in terms of the
\textit{Optimal Sub-Pattern Assignment} (OSPA) metric (with Euclidean distance $c=600 [m]$, and cutoff parameter $p=1$) \cite{SVVospa2008}.

The target state is defined as $x_k=[p_{x,k},\dot{p}_{x,k},p_{y,k},\dot{p}_{y,k}]^{\top}$, where $(p_{x,k},p_{y,k})$ and $(\dot{p}_{x,k},\dot{p}_{y,k})$ are the Cartesian coordinates of position and velocity, respectively. The state transition density, for each target, is assumed to be linear Gaussian, i.e.,
\begin{align}
\nonumber
\label{F}
{f(x_k|x_{k-1})} = {\cal G}(x_k;Ax_{k-1},Q),
\end{align}
with 
\begin{align}
\nonumber
A= \left[ {\begin{array}{*{20}{c}}
	1&T_s\\
	0&1
	\end{array}} \right]\otimes I_2,
Q=\sigma_w^2\left[ {\begin{array}{*{20}{c}}
	\textstyle{T_s^4 \over 4}& \textstyle{T_s^3 \over 2}\\
	\textstyle{T_s^3 \over 2}&T_s^2
	\end{array}} \right] \otimes I_2,
\end{align}
where $I_n$ denotes the $n\times n$ identity matrix and $T_s$ is the sampling interval. 
$A$ and $Q$ are the state transition matrix and process noise covariance with $\sigma_w=5[ms^{-1}]$, respectively.

Each node deployed in the sensor network is able to provide both \emph{Time-of-Arrival} (TOA) and \emph{Direction-of-Arrival} (DOA) measurements, which are characterized by the following observation model:
\begin{align}
\nonumber
z_k^i = \left[ {\begin{array}{*{20}{c}}
	{\text{atan2}({p_{x,k}} - p_x^i,{p_{y,k}} - p_y^i)}\\
	{\sqrt {{{({p_{x,k}} - p_x^i)}^2} + {{({p_{y,k}} - p_y^i)}^2}} }
	\end{array}} \right] + {\varepsilon _k},
\end{align}
where $(p_x^i,p_y^i)$ denotes the known position of node $i$. The covariance of the measurement noises is
\begin{align}
\nonumber
{R} = diag({[\sigma _\theta ^2,\sigma _r^2]^{\top}})
\end{align}
with $\sigma_r =5[m]$ and $\sigma_\theta = 1[^o]$. 

Clutter, for each node, is modeled as a Poisson RFS with an intensity of $\lambda_c=15$, which means that there are, on average, $15$ clutter points per scan. The survival probability for each target is $p_S=0.95$.
\begin{figure}[tb]
	\centering
	\includegraphics[width=1.0\columnwidth,draft=false]{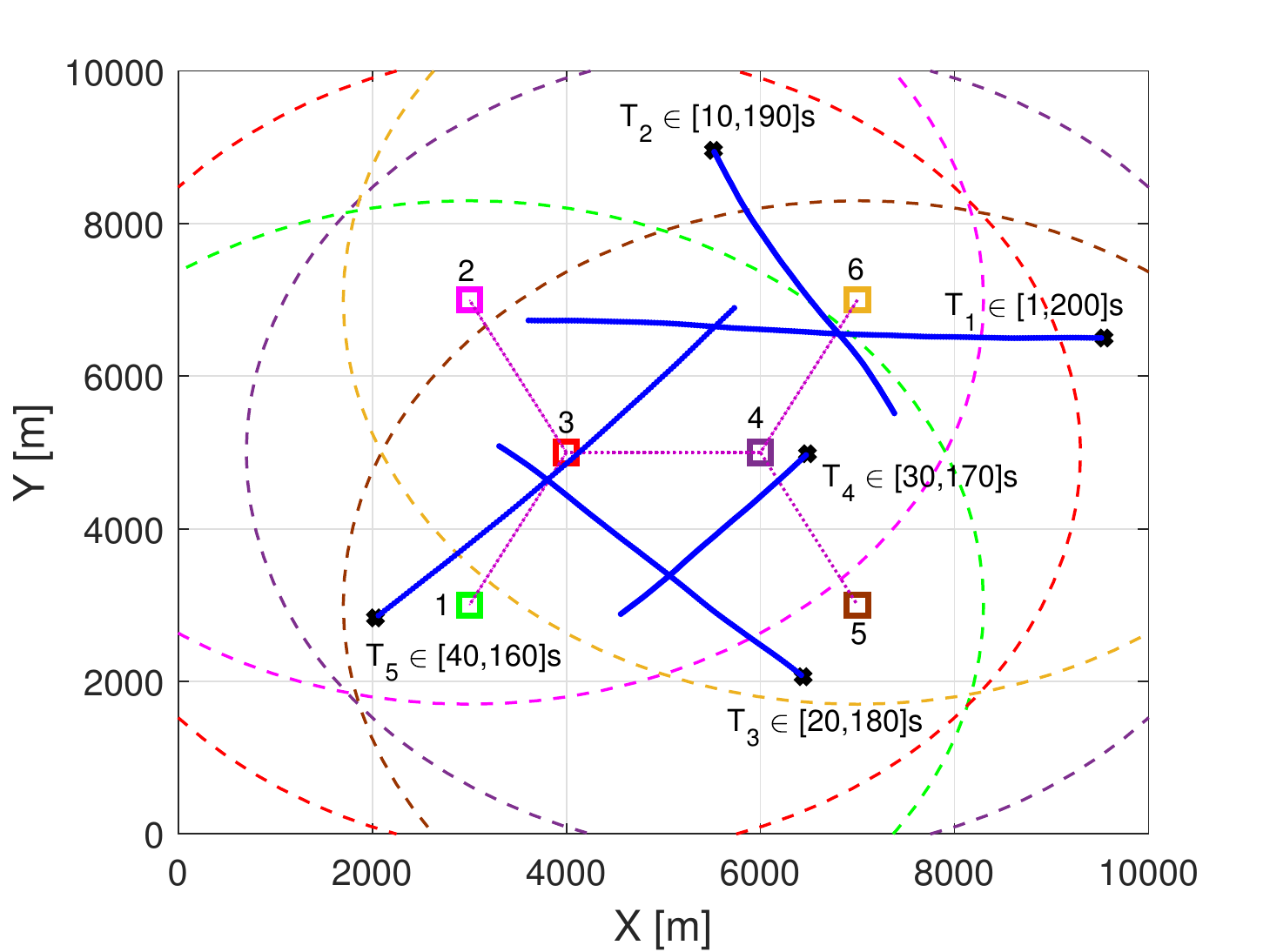}
	\caption{Target trajectories and sensor network considered in the simulated scenario, where the circles with different colors represent different nodes and the dashed lines of the same colors depict the corresponding FoVs. $`\square'$ is the position of node and $`\times'$ denotes the start point for each trajectory.}
	\label{fig:track}
\end{figure}
Each node has a limited FoV as depicted in Fig. \ref{fig:track}. The probability of detection within the FoV is considered constant equal to $p_{D,0} \in (0,1)$, while it is zero outside the FoV. Thus, the detection probability for any node $i$ is modeled as follows.
\begin{align}
\nonumber
{p_D^i}(x) = p_{D,0} \, {1_{{{\cal S}^i}}}(x) \, .
\end{align}

Further, the adaptive birth model based on received measurements is adopted as in \cite{RCVVadaptive2012}. 
Each measurement is regarded as a potentially newborn target with
\begin{align}
\nonumber
m_{\gamma,k}(1) &= z_k(2) ~ \sin(z_k(1)),\\
\nonumber
m_{\gamma,k}(3) &= z_k(2) ~ \cos(z_k(1)),
\end{align}
and $m_{\gamma,k}(2)=m_{\gamma,k}(4)=0$, ${P_b} = \text{diag}{({[50,20,50,20]^{\top}})^2}$. 
The average number of newborn targets per frame is taken as  $0.15$. Further, the cardinality distribution of newborn targets is assumed Poisson. 
The maximum number of Gaussian components is set to $40$ and the maximum number of targets to $20$.
A consensus-based distributed fusion architecture is considered wherein each sensor node fuses its local posterior with
those of its neighbors. More specifically, in each sampling interval, information exchange and fusion are repeated 3 times,
corresponding to 3 consensus steps.

To assess the performance of the proposed distributed fusion with multiple FoVs, 100 Monte Carlo (MC) runs are performed on the same trajectories but with independently generated measurements (including clutter and target-generated measurements) for each trial.

\subsection{Scenario 1}
\begin{figure}[tb]
	\centering
	\includegraphics[width=1.0\columnwidth,draft=false]{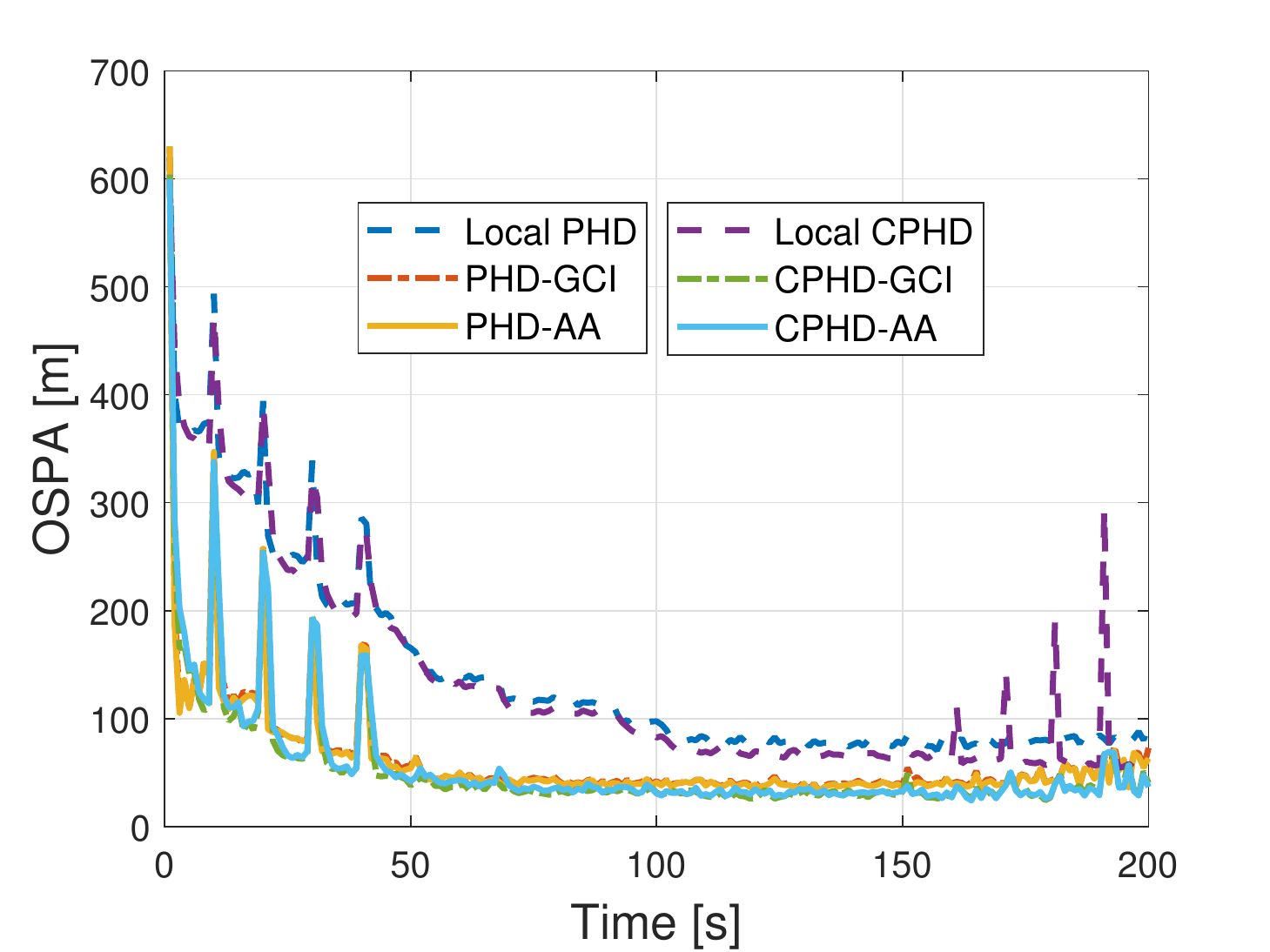}
	\caption{Comparison of OSPA errors between local PHD, local CPHD, PHD-GCI, PHD-AA, CPHD-GCI, and CPHD-AA under high detection probability.}
	\label{fig:ospa95}
\end{figure}

\begin{figure}[tb]
	\centering
	\includegraphics[width=1.0\columnwidth,draft=false]{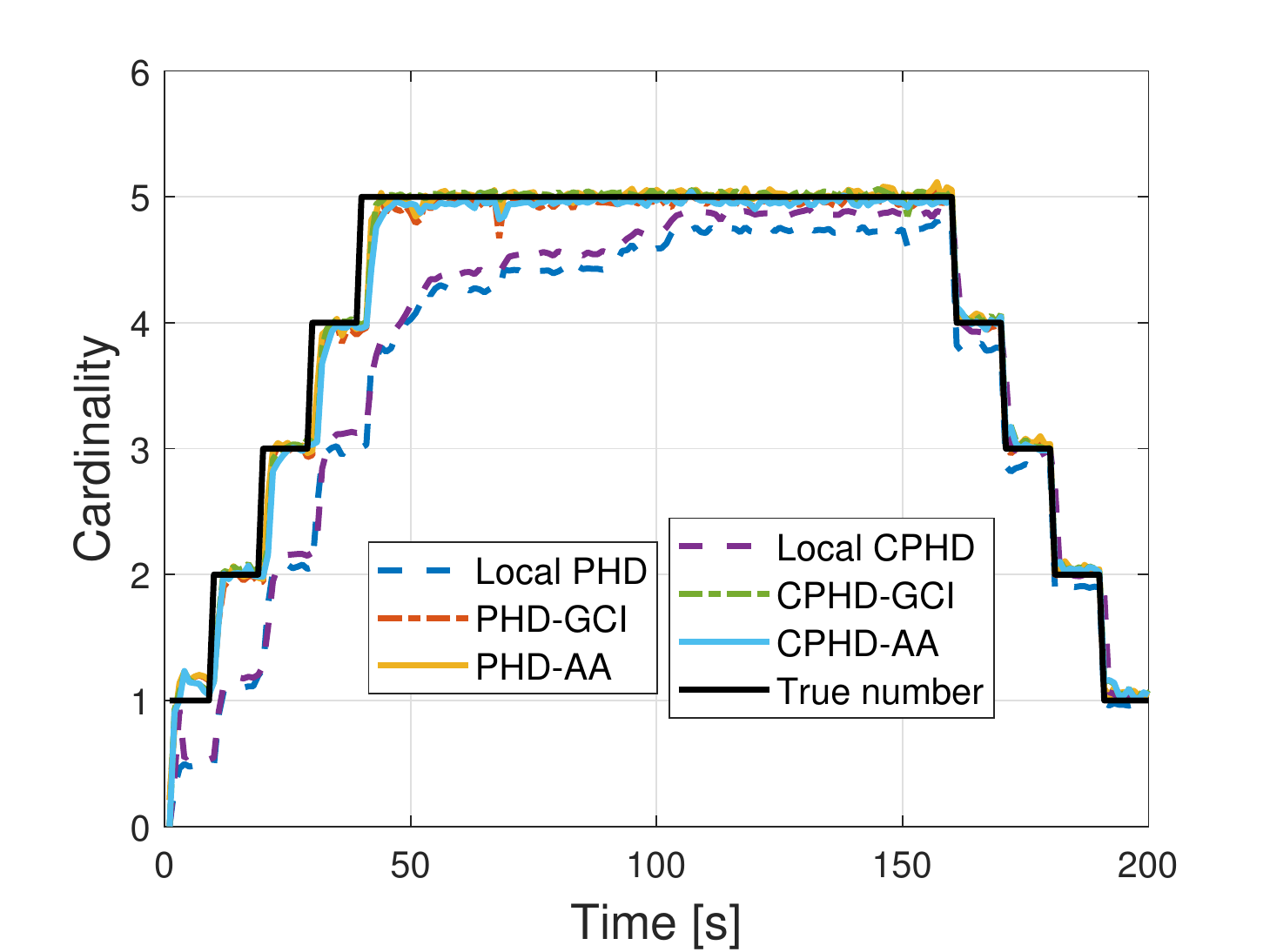}
	\caption{Comparison of cardinality estimates between local PHD, local CPHD, PHD-GCI, PHD-AA, CPHD-GCI, and CPHD-AA under high detection probability.}
	\label{fig:card95}
\end{figure}

Filtering with high detection probability ($p_{D,0}=0.95$) is considered here. The comparisons of OSPA errors and cardinality estimates are respectively shown in Figs. \ref{fig:ospa95}
and \ref{fig:card95}. For both PHD and CPHD filters, the performance of GCI and AA fusion are almost the same. Moreover, the performance of CPHD-based fusion is better than PHD-based one, even if  the gap is not very large. It can also be seen that, after the transient, CPHD-GCI performs slightly better than CPHD-AA. 
Moreover, the filtering performance of single node PHD/CPHD is very poor, especially in the initial stage of filtering. This is because targets do not fully enter the cFoV of all nodes. 

On the other hand, the cardinality comparison shown in Fig. \ref{fig:card95} shows that the target number estimates of local PHD/CPHD filters are very poor before all targets enter the cFoV. It can also be noticed that all fusion approaches provide good results and much better than local PHD/CPHD filters. 



\begin{figure}[tb]
	\centering
	\includegraphics[width=1.0\columnwidth,draft=false]{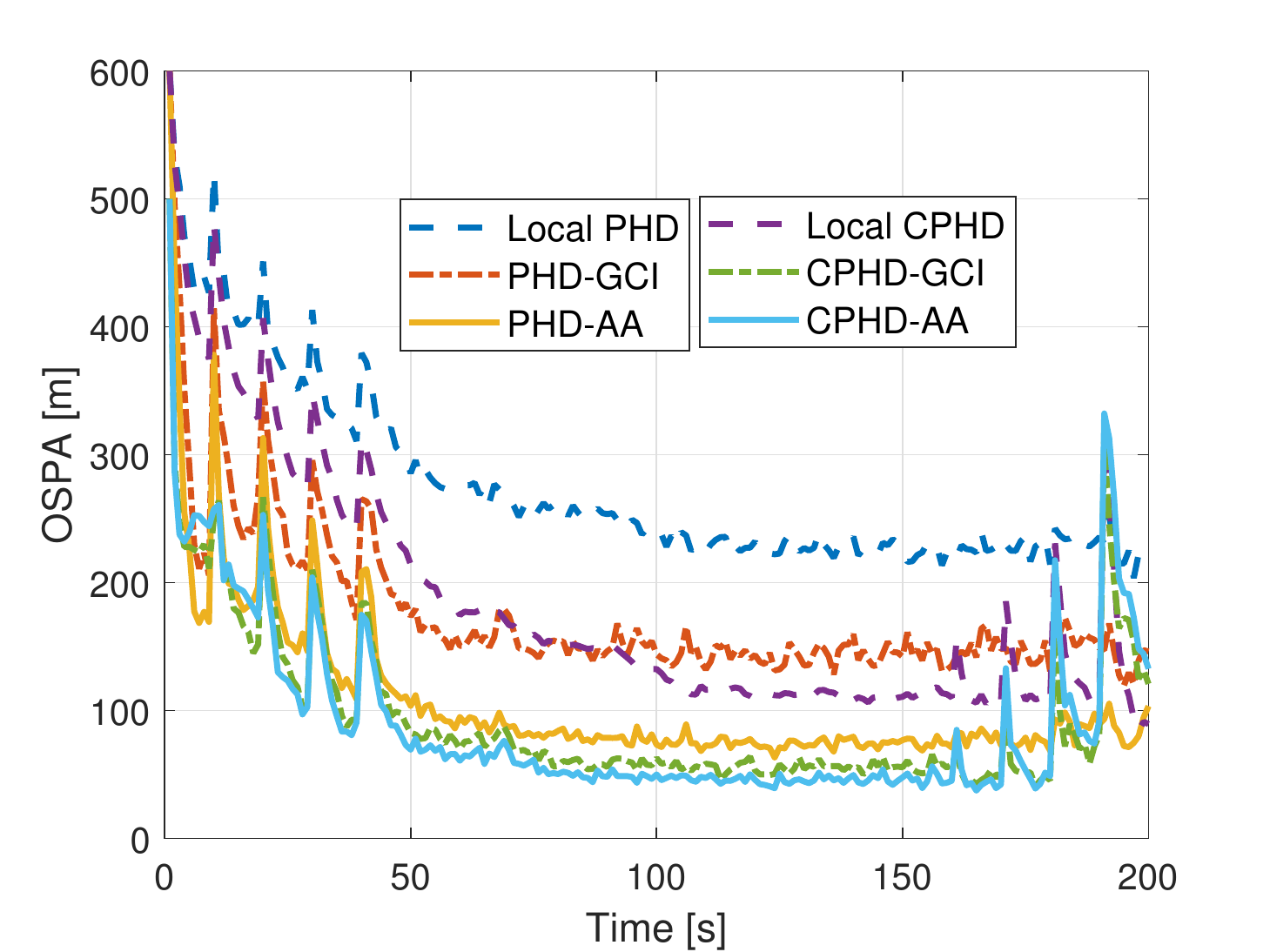}
	\caption{Comparison of OSPA errors between local PHD, local CPHD, PHD-GCI, PHD-AA, CPHD-GCI, and CPHD-AA under low detection probability.}
	\label{fig:ospa65}
\end{figure}

\subsection{Scenario 2}
\begin{figure}[tb]
	\centering
	\includegraphics[width=1.0\columnwidth,draft=false]{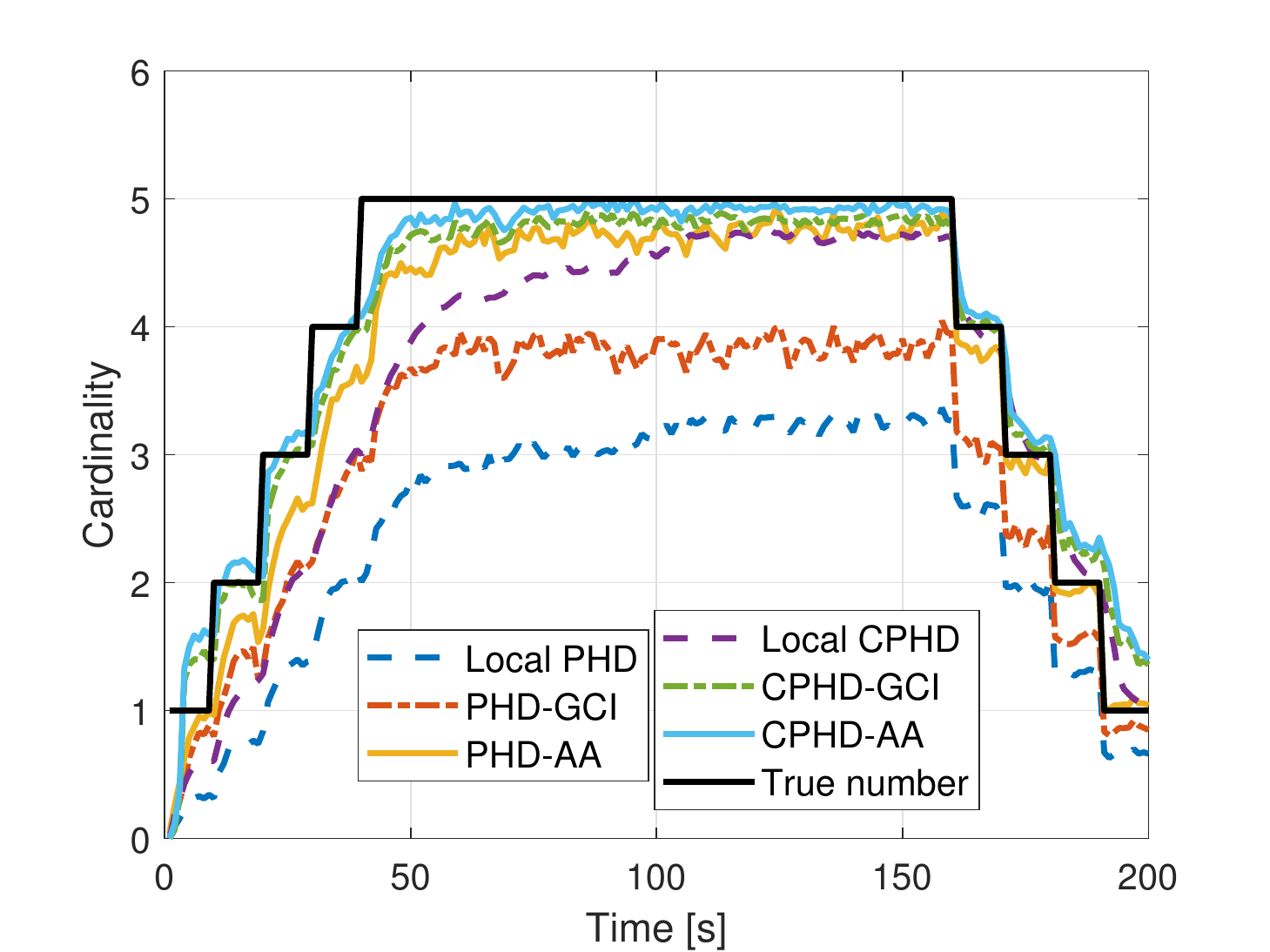}
	\caption{Comparison of cardinality estimates between local PHD, local CPHD, PHD-GCI, PHD-AA, CPHD-GCI, and CPHD-AA under low detection probability.}
	\label{fig:card65}
\end{figure}

In this case, lower detection probability is considered, i.e. $p_{D,0}=0.65$. OSPA errors and cardinality estimates are respectively given in Figs. \ref{fig:ospa65} and \ref{fig:card65}. It is evident that CPHD-AA and CPHD-GCI outperform PHD-AA and PHD-GCI, respectively. 
Meanwhile, it can be found that PHD-based fusion is more sensitive to misdetection than CPHD-based one.
It also turns out that, in this setting, CPHD-AA fusion provides better performance than CPHD-GCI fusion. 
This is due to the multiplicative nature of the GCI fusion rule by which any missed target or existing target with an extremely small weight in a local CPHD filter will reduce the weight of this target in the other nodes, thus reducing the possibility of being detected. 
This problem is also reflected by the
cardinality underestimation of CPHD-GCI fusion shown in Fig. \ref{fig:card65}.

\begin{figure}[tb]
	\centering
	\includegraphics[width=1.0\columnwidth,draft=false]{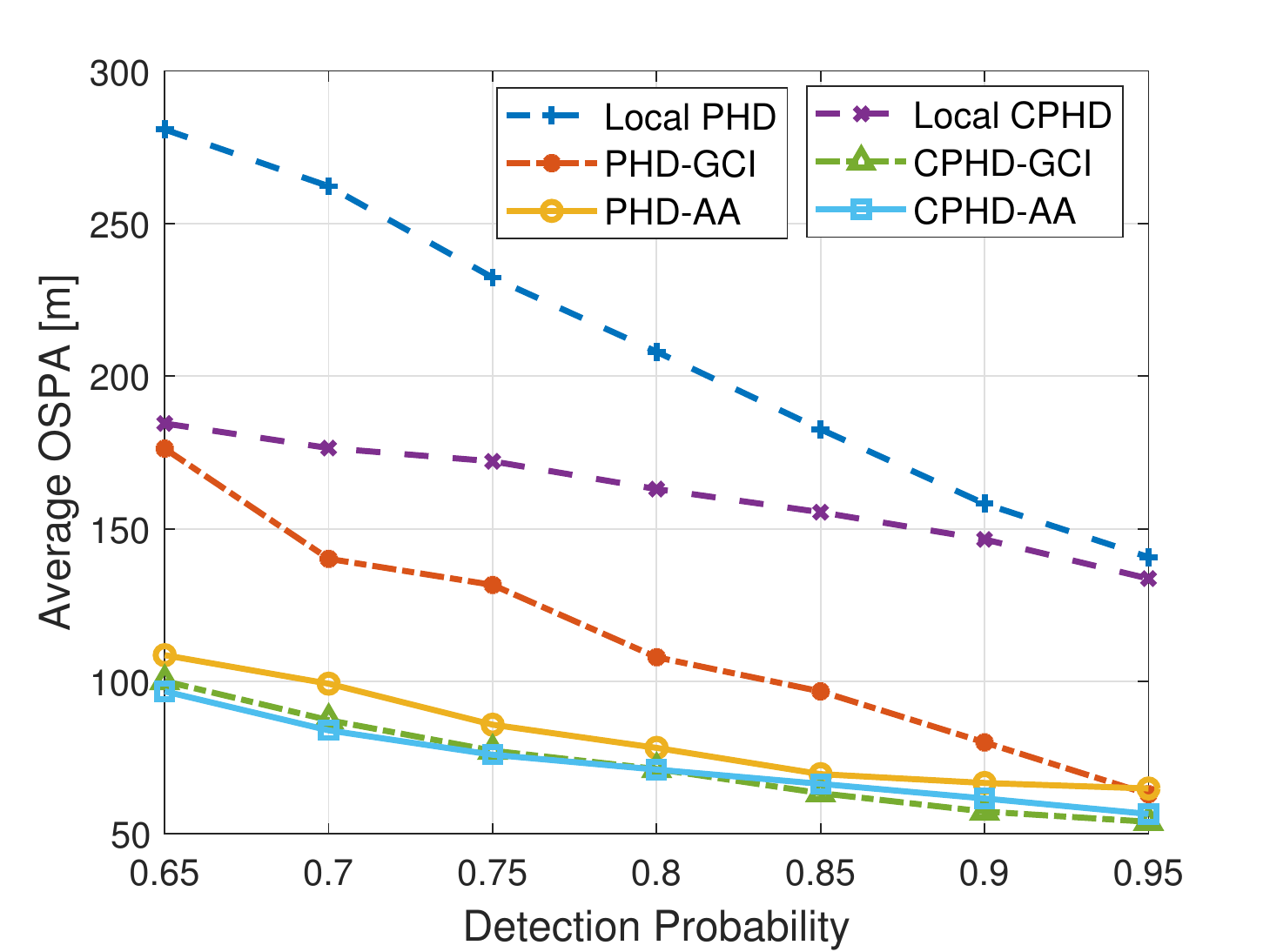}
	\caption{Comparison of average OSPA errors between local PHD, local CPHD, PHD-GCI, PHD-AA, CPHD-GCI, and CPHD-AA under different detection probabilities.}
	\label{fig:comp}
\end{figure}

Next, we test the average OSPA errors under different detection probabilities and the same clutter rate. 
The number of consensus steps of all fusion approaches is set to $L=3$. 
The corresponding result is illustrated in Fig. \ref{fig:comp}. 
It can be seen that CPHD-based fusion is always better than PHD-based fusion and that PHD-GCI fusion is more sensitive to misdetection than CPHD-GCI fusion. 
Furthermore, CPHD-GCI fusion outperforms CPHD-AA fusion under high detection probability while the converse is true under low detection probability. 
Summarizing, CPHD-AA fusion has higher tolerance to misdetection than CPHD-GCI.

\section{Conclusions}

In this paper, we have proposed a distributed multi-target tracking approach based on CPHD filtering under multiple fields-of-view (FoVs). 
First, it has been pointed out how standard fusion is infeasible when the fusing nodes have different FoVs, and it has also been shown that the seemingly reasonable strategy of fusing
information within the common FoV while copying information from the exclusive FoVs is non robust with respect to unavoidable uncertainties on the FoVs.
Next, we have presented a clustering-based approach to decompose the intensity functions to be fused  into multiple clusters.
By approximating the target IID cluster process as multi-Bernoulli RFS, the cardinality distribution of each cluster can be re-constructed. Then, we can perform fusion for all clusters in parallel, exploiting either GCI or AA rules. Finally, all clusters are merged into an IID cluster process.
Moreover, an interpretation of the proposed clustering-based fusion strategy in terms of decomposition of the state space has
been provided, and an analysis of the approximation errors implied by the considered decomposition procedure has been carried out.
Finally, simulation experiments have been provided to demonstrate the effectiveness of the proposed approach.  

\section*{Acknowledgement}
We would like to thank the support of the Scholarship of China Scholarship Council (CSC).

\begin{appendices}

\section{Proofs}

{\em Proof of Lemma 1:} Let $(i,p)$ and $(j,q)$ belong to different clusters.
Then, we have
\begin{equation}\label{eq:rho}
(m_p^i -m_q^j)^\top (P_p^i + P_q^j)^{-1}  (m_p^i -m_q^j) > \rho. 
\end{equation}
Consider now the two ellipsoids ${\cal E}^i_p (\delta)$ and ${\cal E}^j_q (\delta)$ and suppose that they are not disjoint, i.e.
there exists $x \in \mathbb X$ such that the following two conditions hold
\begin{align}
(x-m_p^i)^\top (P_p^i)^{-1}  (x-m_p^i)  \le \delta,\\
(x -m_q^j)^\top (P_q^j)^{-1}  (x -m_q^j) \le \delta. 
\end{align}
In turn, these conditions imply that
\begin{align}
(x-m_p^i)^\top (P_p^i + P_p^j)^{-1}  (x-m_p^i)  \le \delta,\\
(x -m_q^j)^\top (P_p^i + P_q^j)^{-1}  (x -m_q^j) \le \delta,
\end{align}
because $P_p^i$ and $P_q^j$ are positive definite.
Then, we can bound the corrected Mahalanobis distance as
\begin{align}
&(m_p^i -m_q^j)^\top (P_p^i + P_q^j)^{-1}  (m_p^i -m_q^j) \nonumber \\
&\quad = (m_p^i -x + x - m_q^j)^\top (P_p^i + P_q^j)^{-1}  (m_p^i- x + x -m_q^j) \nonumber \\
&\quad \le 2  (m_p^i -x)^\top (P_p^i + P_q^j)^{-1}  (m_p^i- x) \nonumber \\
&\quad \quad +  2 ( x - m_q^j)^\top (P_p^i + P_q^j)^{-1}  (x -m_q^j) \nonumber \\
&\quad \le 4 \delta.
\end{align}
Clearly, this bound is not in contradiction with (\ref{eq:rho}) only if $4 \delta > \rho$. This means that when
$\delta \le \rho/4$ the two  ellipsoids ${\cal E}^i_p (\delta)$ and ${\cal E}^j_q (\delta)$ have to be disjoint or otherwise
we get into a contradiction.  \qed

{\em Proof of Theorem 1:} Recalling (\ref{eq:error}) and applying the triangular inequality, we obtain
\begin{align}
\left \| v^i_g (x) - \hat v^i_g (x) \right \|_1 \le &  \sum_{p: \, (i,p) \in {\cal C}_g } \alpha_p^i \, 
\left \| \left [ 1- 1_{\mathbb X_g} (x)\right ] \, {\cal G}^i_p (x) \right \|_1
\nonumber \\
& + \sum_{p: \, (i,p) \notin {\cal C}_g }   \alpha_p^i \, \left \| 1_{\mathbb X_g} (x)  \, {\cal G}^i_p (x) \right \|_1.
\end{align}
Consider first a generic element of the first summation.
Since $(i,p) \in {\cal C}_g \implies {\cal E}^i_p (\delta) \subseteq \mathbb X_g$, we have
\begin{align}
\left \| \left [ 1- 1_{\mathbb X_g} (x)\right ] \, {\cal G}^i_p (x) \right \|_1 
& = \int_{\mathbb X} \left [ 1- 1_{\mathbb X_g} (x)\right ] \, {\cal G}^i_p (x) \,dx \\
&= 1 - \int_{\mathbb X_g} {\cal G}^i_p (x) \,dx \\
& \le 1- \int_{ {\cal E}^i_p (\delta)}  {\cal G}^i_p (x) \,dx \\
& = 1- F(\delta , \dim (x)).
\end{align}
Consider now a generic element of the second summation.
Since $(i,p) \notin {\cal C}_g \implies {\cal E}^i_p (\delta) \cap \mathbb X_g = \emptyset$, we have
\begin{align}
\left \| 1_{\mathbb X_g} (x) \, {\cal G}^i_p (x) \right \|_1 
& = \int_{\mathbb X_g} {\cal G}^i_p (x) \,dx \\
&= 1 - \int_{\mathbb X \setminus \mathbb X_g} {\cal G}^i_p (x) \,dx \\
& \le 1- \int_{ {\cal E}^i_p (\delta)}  {\cal G}^i_p (x) \,dx \\
& = 1- F(\delta , \dim (x)).
\end{align}
Combining the obtained bounds, we get
\begin{align}
\left \| v^i_g (x) - \hat v^i_g (x) \right \|_1 \le & \sum_{p=1}^{J^i}  \alpha_p^i \, \left [ 1- F(\delta , \dim (x)) \right ]
\end{align}
from which the theorem statement follows. \qed

\end{appendices}



\end{document}